\documentclass[twocolumn,showpacs,preprintnumbers,amsmath,amssymb]{revtex4}
\usepackage{tabularx,graphicx}\begin{document}
\newcommand{\beq}{\begin{equation}}
\newcommand{\eeq}{\end{equation}}
\newcommand{\beqn}{\begin{eqnarray}}
\newcommand{\eeqn}{\end{eqnarray}}
\newcommand{\bmath}{\begin{subequations}}
\newcommand{\emath}{\end{subequations}}
\title{Spin-split states in aromatic molecules and superconductors}
\author{J. E. Hirsch }
\address{Department of Physics, University of California, San Diego,
La Jolla, CA 92093-0319}

\begin{abstract} 
A state where spin currents exist in the absence of external fields has recently been proposed to describe the superconducting state of metals. It is proposed here that such a state also
describes the ground state of aromatic molecules. 
 It is argued that this point of view provides  a more natural explanation for the large diamagnetic susceptibilities and NMR
shifts observed in these molecules than the conventional viewpoint, and it provides a unified description of aromatic molecules and superconductors as sought by
F. London. A six-atom ring model is solved by exact diagonalization and parameters in the model where   a ground state spin current exists
are found. We suggest that this physics plays a key role in  biological matter.
   \end{abstract}
\maketitle

\section{introduction}
The large anisotropic diamagnetic susceptibility exhibited by aromatic molecules
and the associated NMR frequency shifts at nearby atoms are generally interpreted as arising from delocalized ring currents in the molecular rings\cite{aromaticity}. This
interpretation relies on molecular orbital theory (Huckel theory) of $\pi$ electrons and assumes electrons are non-interacting\cite{pauling,haddon}.
On the other hand, calculations using valence bond theory taking electron-electron interaction into account predict  that the $\pi$ electrons in aromatic rings are localized\cite{vb}.
These electrons move in a one-dimensional half-filled band, where the effect of electron-electron interactions should be strong;
electron-electron interactions in a one-dimensional half-filled band can  lead to localization of electrons (Mott insulating state) which
would suppress the diamagnetic response. Within valence bond theory the aromatic character is proposed to result from the coupling of electron spins\cite{vb}.

F. London suggested long ago that the diamagnetic currents in aromatic rings are analogous to supercurrents in superconductors\cite{london}.
At that time superconductivity in metals was not understood from a microscopic point of view. 
After the advent of BCS theory the possible connection between ring currents in aromatic molecules and superconducting currents has been discussed by
W. Little\cite{little} in connection with a proposed excitonic pairing mechanism. However, to date no compound has been identified that
would
exhibit Little's excitonic superconductivity. 

The present author suggested in 1990 that the large diamagnetic response of aromatic ring molecules may originate in the existence of 
a spin current in the ground state of such molecules in the absence of applied fields\cite{ss}. Thus, an applied magnetic field would not set electrons into motion but rather
slow down the component of the pre-existent spin current in one direction and speed up the spin current in the opposite direction. 
 A heuristic model
for this physics was proposed and it was suggested that it provides a natural explanation for various experimental observations.
At that time we did not make a connection between this physics and superconductivity.

However our recent work on the theory of hole superconductivity\cite{holesc}  led   to the conclusion that metals expel negative charge from the interior towards the surface in the
transition to the superconducting state\cite{chargeexp}. This phenomenon (which also provides a `dynamical' explanation of the Meissner effect\cite{missing}) leads to the prediction
that a  
a macroscopic spin current exists near the surface of superconductors in the absence of applied external fields (Spin Meissner effect)\cite{spincurrent,sm}.  
Naturally this raises the question  whether the spin currents in aromatic molecules proposed in ref.\cite{ss} are related to the 
spin currents in superconductors recently predicted\cite{sm}. In this paper we propose that in fact such a deep connection exists, validating the early
intuition of F. London. A state qualitatively different from the Huckel\cite{pauling,haddon} and valence bond\cite{vb} states, carrying a spin current around the ring, is proposed to describe the ground state of aromatic molecules.

\section{spin current and $\pi$ flux in superconductors}
We have recently proposed that the Meissner effect in superconductors can be understood
`dynamically' if electrons expand their orbits to mesoscopic orbits of radius $2\lambda_L$ in the
transition to superconductivity, with $\lambda_L$ the London penetration depth\cite{sm}. In the absence
of applied magnetic field this orbit expansion results, through the spin-orbit interaction of the electron
magnetic moment $\mu_B=e\hbar/2m_ec$  with
the positive ionic background $|e|n_s$ ($n_s=$superfluid density), in a spin current with speed\cite{sm}
\beq
v_\sigma^0=\frac{\hbar}{4m_e\lambda_L}
\eeq
and opposite direction for electrons with opposite spin.

The angular momentum of electrons in orbits of radius $2\lambda_L$ with speed Eq. (1) is
\beq
L=m_e v_\sigma^0 (2\lambda_L)=\frac{\hbar}{2}
\eeq
This remarkable result indicates that electrons in superconductors have an orbital `spin' analogous
to the intrinsic electron spin. A spinning electron can be visualized as a charge orbiting at
speed $c$ in a circle of quantum electron radius
\beq
r_q=\frac{\hbar}{2m_e c}
\eeq
and the electron in the superconductor can be interpreted as an `amplified image'
of the spinning electron, with amplification factor $2\lambda_L/r_q$\cite{holeelec4}. 

An angular momentum $\hbar/2$ corresponds to a wavefunction $\Psi=|\Psi|e^{i\theta}$ depending on the azimuthal angle  $\varphi$ as
\beq
\Psi \propto e^{i\theta(\varphi)} = e^{\pm i\varphi/2}
\eeq
 so that the electron acquires a phase $\pi$ in traversing a full orbit. This in turn can be understood as
 arising from a `flux' $\phi_0$ inside the orbit giving rise to the phase factor
 \beq
 e^{\frac{ie}{\hbar c}\oint \vec{A}_\sigma \cdot \vec{dl}}=   e^{\frac{ie}{\hbar c}\sigma\phi_0 }=e^{\pm i\pi}
\eeq
with $\phi_0=\frac{hc}{2e}$ the flux quantum and $A_\sigma$ the
spin-orbit vector potential\cite{electrospin}.

Thus, the development of  the macroscopic spin current predicted by the theory of hole superconductivity
upon the establishment of  macroscopic phase coherence in superconductors  requires
 `closing'\cite{nikulov}  of the wavefunction for each member of the Cooper pair according
to the relation
\beq
\oint \vec{\nabla}\theta_\sigma\cdot\vec{dl}=\pm \pi
\eeq
where $\theta_\sigma$ is the phase of the member of the Cooper pair of spin $\sigma$,
instead of the usual Bohr-Sommerfeld  quantization rule.
Alternatively we can think of   $\theta_\sigma$ as characterizing the phase of the entire condensate of spin $\sigma$ rather than a single member of a Cooper pair,
since the phases of different Cooper pairs are phase locked due to macroscopic phase coherence.
The phase factor $e^{\pm i\pi}=-1$ acquired by one spin component is cancelled by the same factor in the opposite spin component  so that the superconducting wavefunction is single-valued.

As is well known, for an individual spin-$1/2$ spinor, a rotation by $360$ degrees gives rise to a change in sign\cite{merzbacher}, which we can 
also interpret as arising from an enclosed $\pi-$flux.
Thus, a single spinning electron shows the same phase behavior as the Cooper pair member in the mesoscopic orbits 
in the superconductor within the theory
of hole superconductivity,
proposed to describe all superconductors\cite{bcsquestion}. The fact that the same physics
shows up at the subatomic scale $r_q$ (Eq. (3)) and at the mesoscopic scale $2\lambda_L$ suggests 
that it will also show up at  other length scales provided phase coherence exists. In this paper we propose that this
physics shows up at the atomic scale $a_0=\hbar^2/m_e e^2 \sim \sqrt{r_q(2\lambda_L)}$
and in particular manifests itself in the existence of a spin current in the ground state
of aromatic molecules such as benzene.

\section{spin current and $\pi$ flux in benzene}

The aromatic character of benzene is ascribed to the six electrons in the $p\pi$ orbitals oriented perpendicular to the plane of the molecule.
There are two possible orientations for the atomic wave function, as shown schematically in Fig. 1.
   \begin{figure}
 \resizebox{7.0cm}{!}{\includegraphics[width=7cm]{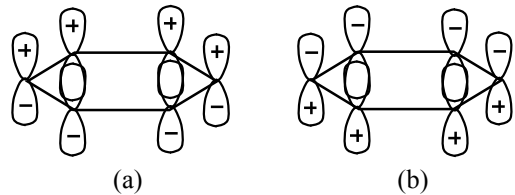}}
 \caption  {The two possible orientations of the carbon $p_z$ orbitals in the benzene molecule. In the conventional point of view only 
 one of them (either one) plays a role.}
 \label{figure2}
 \end{figure}
 In the conventional point of view, these two orientations  are equivalent except for a sign convention which has no physical significance, and either one of them is used. 
 Instead, we propose here that the fact that there are two possible orientations of the $p_z$ orbital as shown in Fig. 1 
 {\it has profound physical significance}.
 
We propose, by analogy with the situation for the intrinsic electron spin\cite{merzbacher}, as well as for the $2\lambda_L$ orbits in the
 superconductor discussed in the previous section, that as an electron goes around once in the aromatic ring, it ends up   in the orbital with
 opposite sign orientation, as shown schematically in Figure 2. That is, the electron  acquires a $\pi $ phase shift, and two rounds are needed to bring
 the electron back to its original state with the same sign. Because for every spin-up electron there is a spin-down electron moving around in
 opposite direction, no sign change occurs for the wavefunction of the system as a whole, just like in the case of the superconductor.
 
    \begin{figure}
 \resizebox{8.5cm}{!}{\includegraphics[width=8cm]{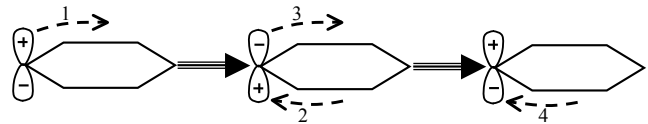}}
 \caption  {In the point of view proposed here, as the electron goes around the benzene ring once, it ends up in the orbital with opposite sign. Two rounds are needed to
 go back to the original state.}
 \label{figure2}
 \end{figure}

     \begin{figure}
 \resizebox{8.5cm}{!}{\includegraphics[width=8cm]{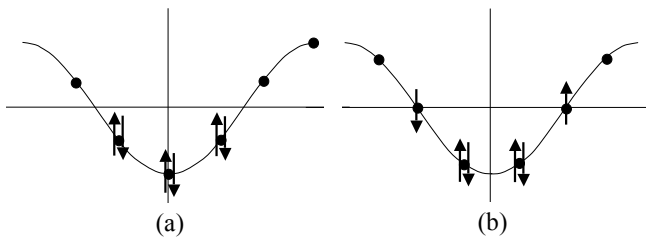}}
 \caption  {In the conventional scenario (a), the  ground state of benzene  is non-degenerate and there is no spin current.
 In the scenario proposed here (b) the (non-interacting)  ground state is four-fold degenerate and can carry spin current.}
 \label{figure2}
 \end{figure}

 To describe this scenario we may take the tight binding Hamiltonian kinetic energy with hopping between nearest neighbors $<ij>$
 \beq
 H_{kin}=-\sum_{<ij>\sigma}t_{ij}(c_{i\sigma}^\dagger c_{j\sigma}+h.c.)
 \eeq
 and take hoppings $t_{ij}=t$ for all bonds except one, where $t_{ij}=-t$, corresponding to `antiperiodic' boundary conditions. 
 Thus, the electrons acquire a phase $\pi$ in going around the ring. The non-interacting spectrum and the occupation of the states for 6 electrons is shown in Fig. 3 
contrasted with  the conventional
 case. In the scenario considered here there are four degenerate ground states
 in the non-interacting case, shown in Fig. 4. Two of them  carry a spin current.
 
 The energy level structure shown in Fig. 4  would arise within the conventional viewpoint if there is an applied magnetic field with flux
 $\phi_0$ within the aromatic ring. Here we propose  that this structure arises spontaneously in benzene in the absence of applied magnetic field.
 It can be thought of as arising from $\pi-f$lux of opposite sign affecting the states of up and down electrons and parallels the 
 behavior predicted for superconductors by the theory of hole superconductivity\cite{sm}.
 
      \begin{figure}
 \resizebox{8.5cm}{!}{\includegraphics[width=8cm]{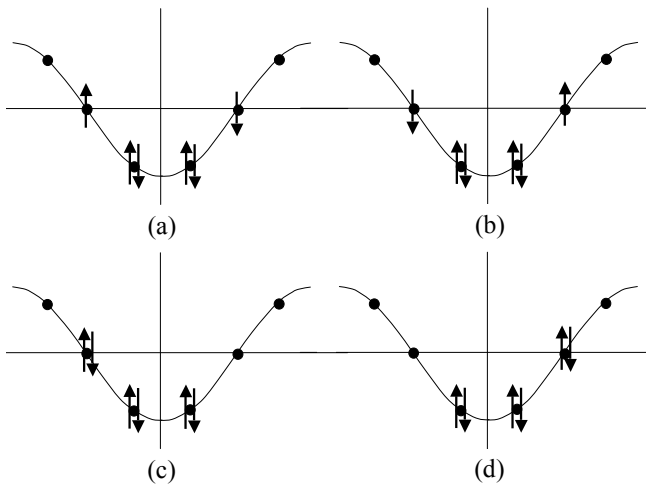}}
 \caption  {  In the scenario proposed here,  the non-interacting ground state is four-fold degenerate as shown in the Figure. Two  of the ground states, 
 (a) and (b),  carry spin current and the other two, (c) and (d), carry charge current.
}
 \label{figure2}
 \end{figure}
 
 Alternatively we argue as follows. Consider a rotation of the coordinate system by an angle $\varphi=\pi/3$ around the $z-$axis chosen perpendicular to the
 plane of the aromatic ring. This is a symmetry operation that leaves the benzene ring invariant. A spinor $\chi$ localized at a given atom in the ring
 is described in the new coordinate system by the spinor\cite{merzbacher}
 \beq
 \chi'=e^{-i\frac{\varphi}{2}\hat{n}\cdot \vec{\sigma}}\chi \equiv e^{-i\theta \sigma_z}\chi    ,
 \eeq
with $\theta=\pi/6$. Six such transformations in sequence bring the coordinate system back to the original one, however the
 spinor changes its sign. We now consider these symmetry operations from an ``active'' rather than a ``passive'' point of view.
 Namely, instead of rotating the observer's coordinate system we rotate our physical system, the spinor. If $\chi$ is an eigenstate of
 $\sigma_z$, it acquires a phase factor $e^{\pm i \pi/6}$ in hopping from a site to a neighboring site, with opposite sign phase factor for 
 both eigenstates of $\sigma_z$.
  
 Adopting this point of view, the kinetic energy operator is, instead of Eq. (7) 
 \bmath
 \beq
 H_{kin}= - \sum_{<ij>\sigma}[ t_{ij}^\sigma c_{i\sigma}^\dagger c_{j\sigma}+h.c.]
 \eeq
 \beq
 t_{ij}^\sigma=te^{i(i-j)\theta_\sigma}
 \eeq
 \beq
 \theta_\sigma=\frac{\pi}{6}\sigma
 \eeq
 \emath
 and the spin quantized in direction perpendicular to the plane of the molecule. The energy spectrum is identical to that shown in Figure 4.
In Eq. (9a),  $<ij>$ indicates nearest neighbor atoms with $i$ to the left of $j$.

 \section{electron-electron interactions}
 
 Can   states that carry a spin current, such as those shown in Fig. 4 (a) and (b), be stable in the presence of strong electron-electron interactions? 
 It may be suspected that in a finite ring electron-electron scattering will destroy the spin current by coupling states with spin current flowing in opposite
 directions\cite{wu}.
 
 We consider the kinetic energies  Eq. (7) or (9) in the presence of on-site and nearest neighbor Coulomb repulsion as well as nearest neighbor
 ferromagnetic exchange
 \beqn
 H&=&H_{kin}+U\sum_i n_{i\uparrow}n_{i\downarrow}+V\sum_{<i.j>\sigma} n_in_j \nonumber \\
 &+&\sum_{<i.j>\sigma \sigma '}  J_{ij}^{\sigma \sigma'}c_{i\sigma}^\dagger c_{j\sigma}c_{j\sigma '}^\dagger c_{i\sigma '}
 \eeqn
 With the kinetic energy Eq. (7) we have simply $J_{ij}^{\sigma \sigma'}=J>0$. This term 
 (ferromagnetic nearest neighbor exchange term) arises from an off-diagonal matrix element of the Coulomb interaction in the tight binding formulation
 ($J$) which is always positive\cite{metallic}. It was proposed in Ref.\cite{ss} that it plays an essential role in stabilizing the spin-split state in aromatic molecules.
 However, in Ref.\cite{ss}, the spin current state envisioned was based on excitations of the conventional non-interacting system Fig. 3(a). We find here that $J$
 also plays an essential role in stabilizing the spin current state based on the scenario of Fig. 4.
 If  we use the kinetic energy Eq. (9), the $J$ interaction   in Eq. (10) takes the form
 \beq
 J_{ij}^{\sigma \sigma'}=J e^{i(i-j)(\theta_\sigma-\theta_{\sigma '})}
 \eeq
 so that it acquires a phase factor for exchange of electrons of opposite spin. The spectrum of the Hamiltonian Eq. (10) is the same for the two forms of the
 kinetic energy Eq. (7) and Eq. (9) and we will use Eq. (9) in what follows.

 We diagonalize the Hamiltonian Eq. (10) exactly for 6 electrons with equal number of up and down spins. It is easy to see that in the presence of interactions
 the ground state cannot carry a spin current if it is non-degenerate, since such a spontaneously broken symmetry state cannot exist in a finite system. So we look for 
 parameters in the Hamiltonian that will give rise to a doubly degenerate or nearly doubly degenerate ground state. 
  Indeed we find that generically such Hamiltonian parameters exist even for the case of strong interactions.
We find them in the neighborhood of  $U\sim 2V$ and they require non-zero values of the exchange interaction $J$.

       \begin{figure}
 \resizebox{8.5cm}{!}{\includegraphics[width=8cm]{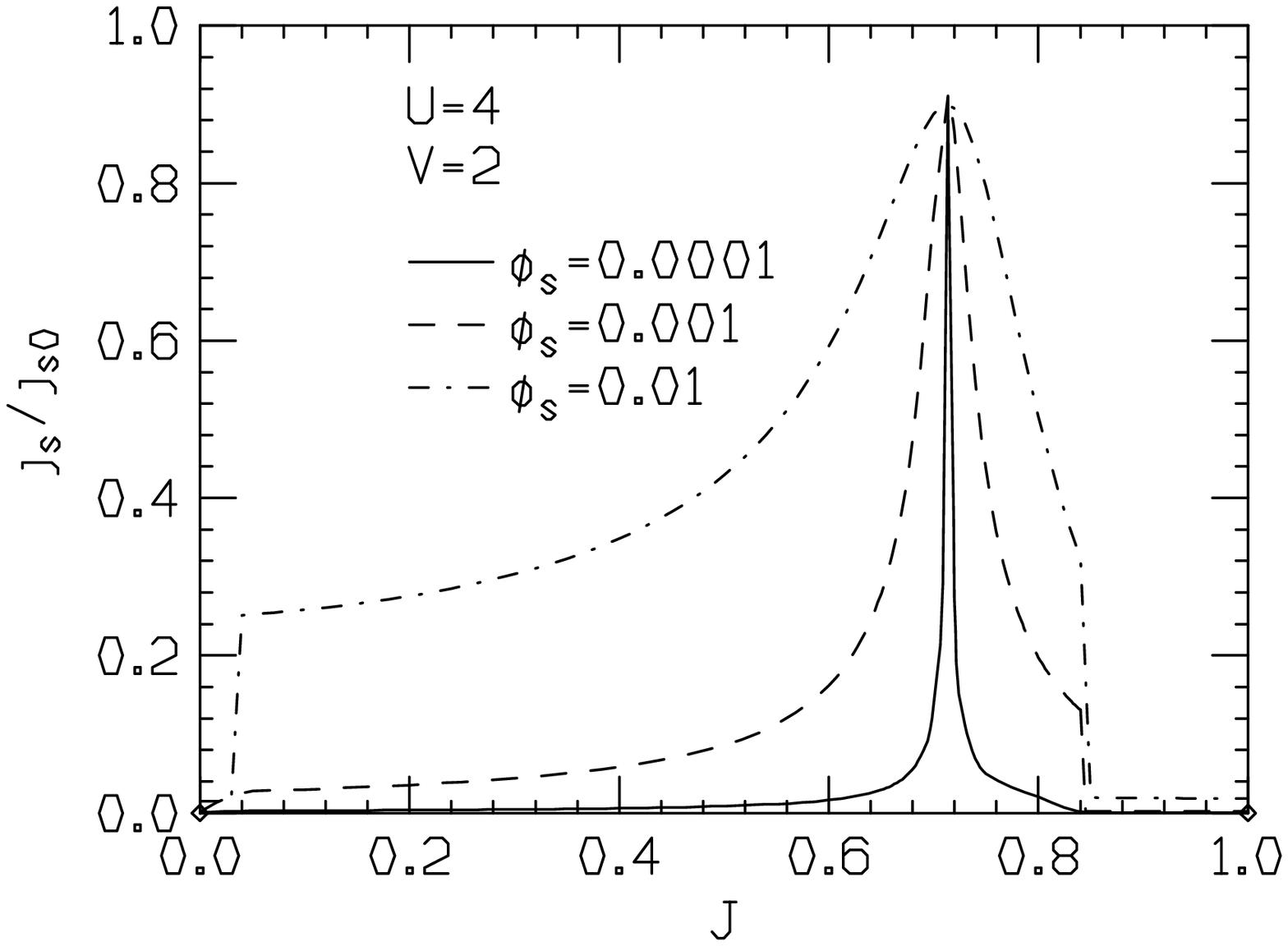}}
 \caption  {  Spin current versus nearest neighbor ferromagnetic exchange interaction strength for three values of the staggered flux $\phi_s$  in Eq. (12).
}
 \label{figure2}
 \end{figure}

       \begin{figure}
 \resizebox{8.5cm}{!}{\includegraphics[width=8cm]{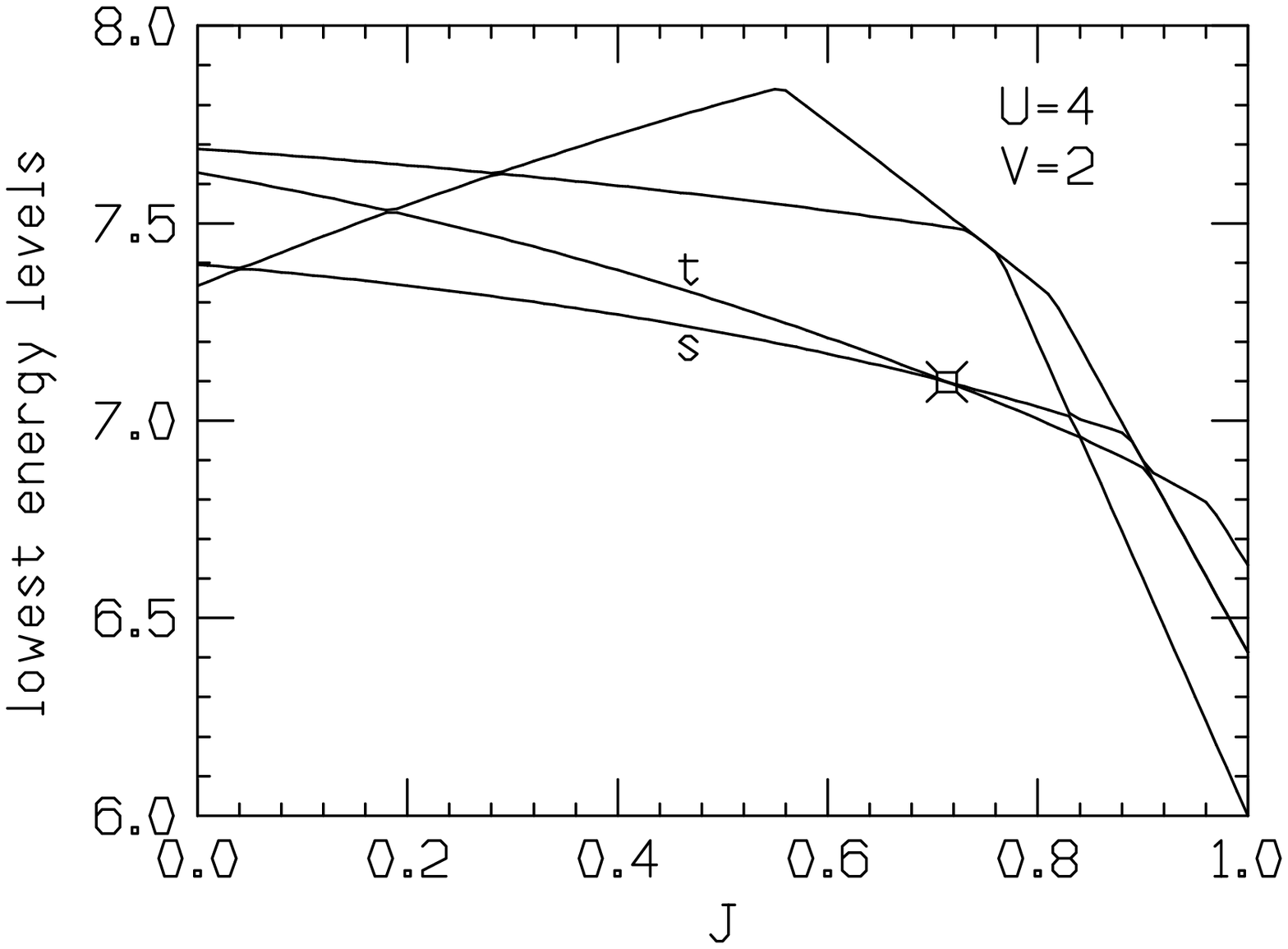}}
 \caption  {  Four lowest energy levels of the Hamiltonian for $U=4$, $V=2$ as function of $J$.
 The ground state switches from singlet ($s$)  to triplet ($t$) at 
 $J\sim 0.714$. At that point (indicated by the symbol in the figure) the degenerate ground state corresponds to the 
 spin-current carrying states of Fig. 4 (a), (b).
}
 \label{figure2}
 \end{figure}
  
       \begin{figure}
 \resizebox{8.5cm}{!}{\includegraphics[width=8cm]{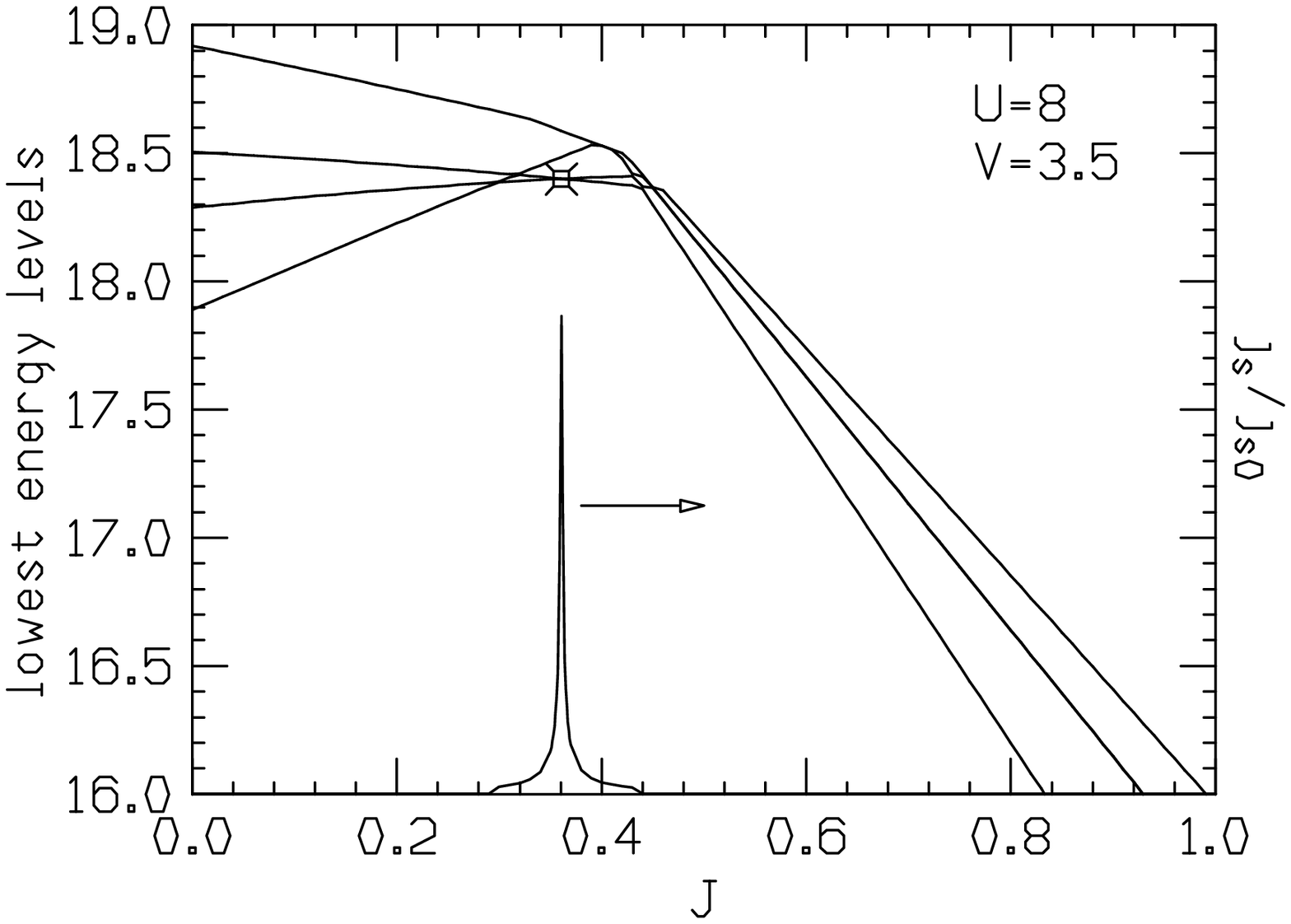}}
 \caption  {  Four lowest energy levels of the Hamiltonian for $U=8$, $V=3.5$ as function of $J$.
 The singlet-triplet crossing point where the spin current is largest is at $J\sim 0.361$.
}
 \label{figure2}
 \end{figure}

       \begin{figure}
 \resizebox{8.5cm}{!}{\includegraphics[width=8cm]{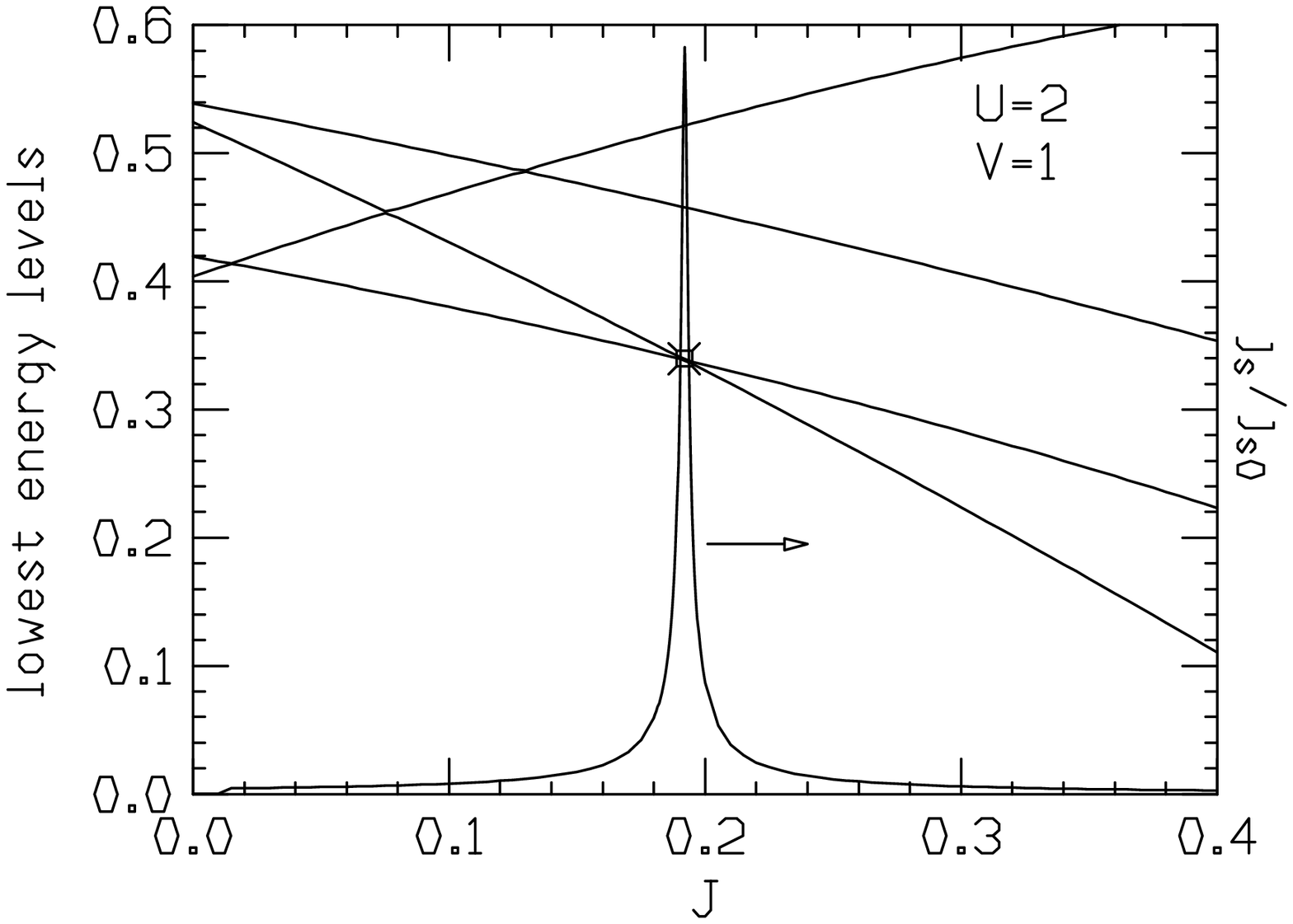}}
 \caption  {  Four lowest energy levels of the Hamiltonian for $U=2$, $V=1$ as function of $J$.
  The singlet-triplet crossing point where the spin current is largest is at $J\sim 0.192$.}
 \label{figure2}
 \end{figure}
 
 Consider for example the case $U=4$, $V=2$ (with $t=1$). To detect a spin current we add to the phases of the spinors a small
staggered flux $\phi_s$, i.e. take
\beq
\theta_\sigma=(\frac{\pi}{6}+\phi_s)\sigma .
\eeq
This arises in the real system from the spin-orbit interaction not included in the Hamiltonian Eq. (10), as discussed in the next section.
Figure 5 shows the ground state spin current 
\beq
j_s=\frac{1}{2}(<\j_\uparrow>-<j_\downarrow>)
\eeq
 with
 \beq
 j_\sigma=-i(t_{ij}^\sigma c_{i\sigma}^\dagger c_{i+1,\sigma}-h.c.)
 \eeq
 versus the nearest neighbor exchange amplitude $J$ for three small values of the staggered flux $\phi_s$. 
 It can be seen that a spin current exists for finite staggered flux and non-zero $J$. As the staggered flux approaches zero, the spin current approaches a 
 $\delta-$function at one value of $J$, $J=0.714$ for this case. This corresponds to the crossing of the two lowest  energy  levels, 
 a singlet and a triplet, as shown in
 Fig. 6. In the limit $\phi_s\rightarrow 0$, spin current will only exists at the precise point where the energy levels cross.

 Similarly we show in Figs. 7 and 8 the four lowest energy levels and the ground state spin current for symmetry-breaking 
 field $\phi_s=0.0001$ for two other sets of $U, V$ values. It can be seen that the behavior is generic.
 
             \begin{figure}
 \resizebox{8.5cm}{!}{\includegraphics[width=8cm]{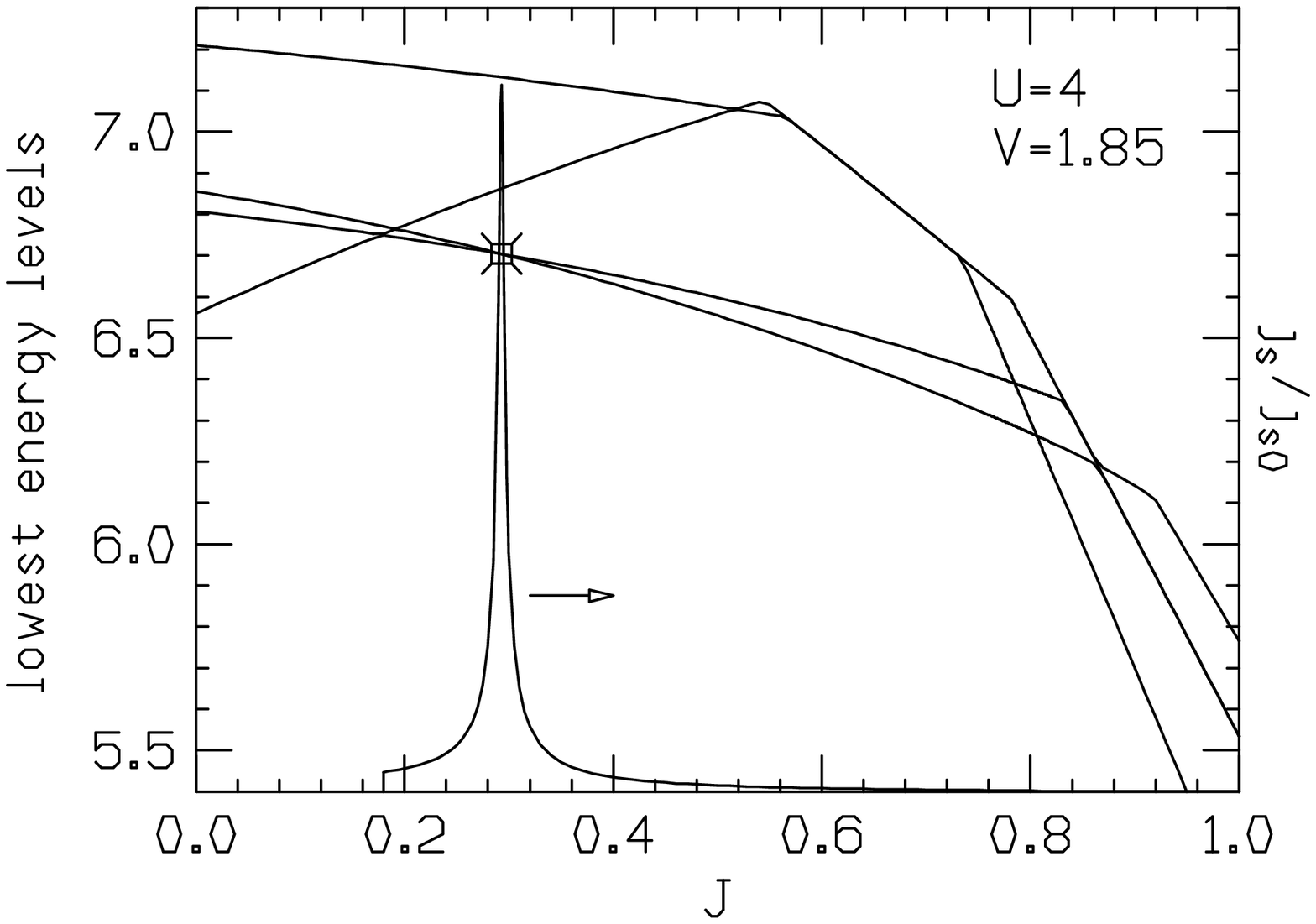}}
 \caption  {  Four lowest energy levels of the Hamiltonian for $U=4$, $V=1.85$ as function of $J$.
 The singlet-triplet crossing point where the spin current is largest is at $J\sim 0.293$.}
 \label{figure2}
 \end{figure}
  Note that the range of parameters in the model where we find this interesting behavior is quite reasonable. For example, first principles
  calculations for benzene yield values $t=2.40eV$, $U=11.26eV$\cite{zoos}. 
  The nearest neighbor repulsion $V$ estimated from the Ohno formula\cite{ohno}
  $V(R)=U/\sqrt{1+R^2U^2/e^4}$ yields $V=7.6eV$ for nearest neighbor distance $R=1.4A$. Hence 
   $U/t=4.7$, $V/t=3.2$. The estimation of these parameters is subject to
  ambiguity, for example  related to the validity of the zero differential overlap approximation\cite{chemphys}. 
  Furthermore the parameter values could be renormalized by inclusion of further neighbor interactions in the model.
  We will investigate these questions in separate work.

             \begin{figure}
 \resizebox{8.5cm}{!}{\includegraphics[width=8cm]{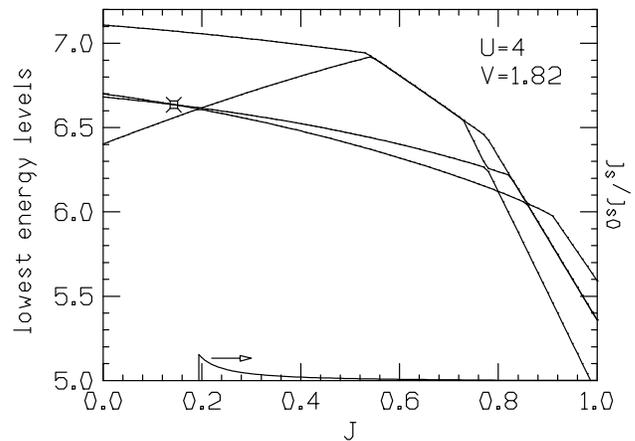}}
 \caption  { Four lowest energy levels of the Hamiltonian for $U=4$, $V=1.82$ as function of $J$.
 The singlet-triplet crossing point is at $J\sim 0.143$ indicated by the symbol in the figure, 
 however another singlet energy level lower in energy exists for
 that $J$ that  does not carry a spin current.}
 \label{figure2}
 \end{figure}

  The range of nearest neighbor interaction values where this behavior occurs is rather restricted.  Fig. 9
  shows the lowest energy levels for $U=4$, $V=1.85$. The crossing point where spin current exists for
  infinitesimal spin-orbit coupling
  has moved from $J=0.714$ down to $J=0.293$. For smaller  $V$ the crossing point for these energy levels moves 
  further to the left  and 
  crosses another energy level,  that becomes the lowest energy state and does not carry spin current,
as shown in Fig. 10 for $U=4$, $V=1.82$. However even in this situation 
if the spin-orbit interaction is finite the ground state and first excited state will carry a spin current for $J$ larger than the crossing
point with the non-spin-current carrying ground state ($J>0.2$ in Fig. 10).

           \begin{figure}
 \resizebox{8.5cm}{!}{\includegraphics[width=8cm]{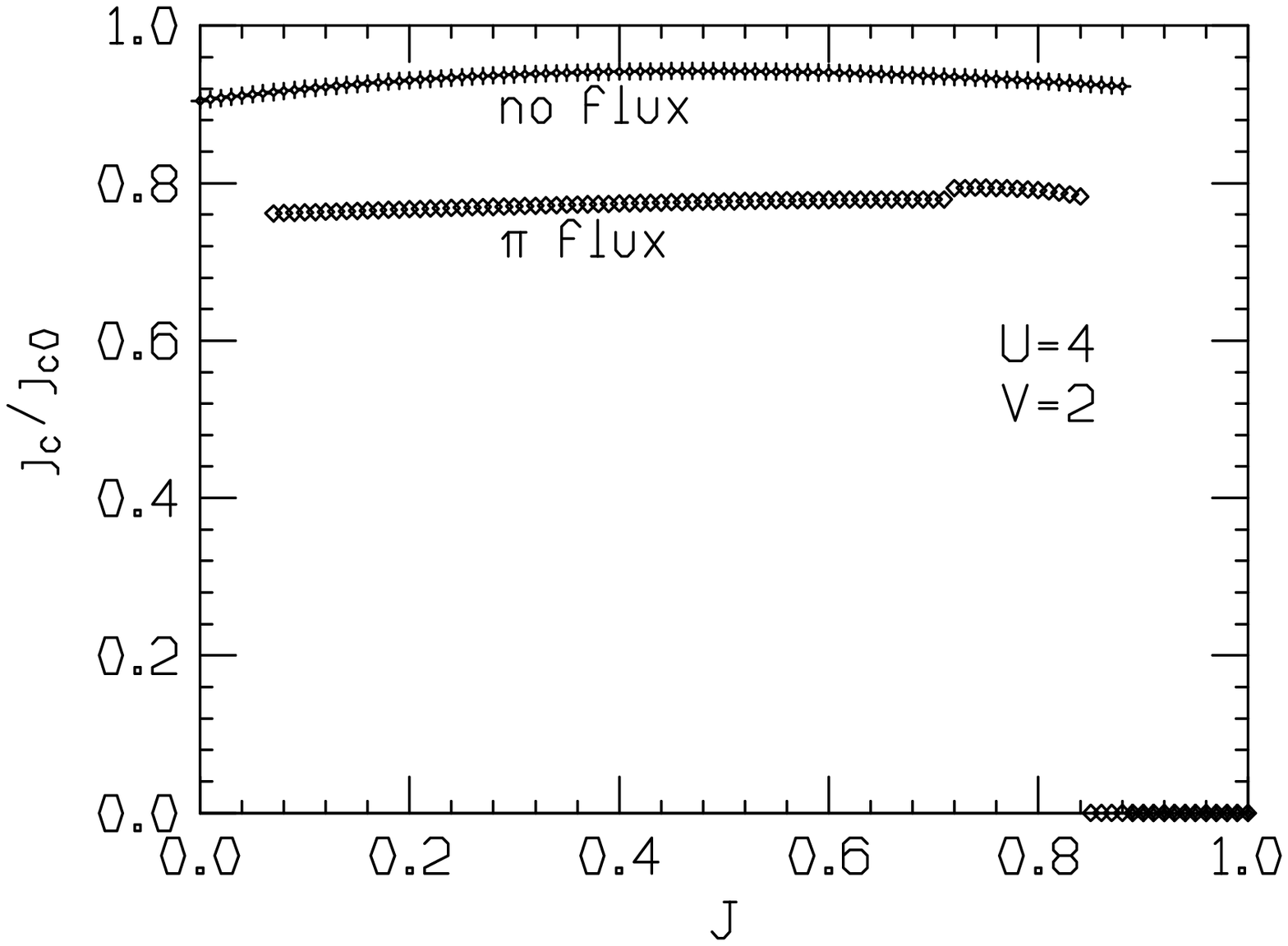}}
 \caption  {  Charge current response to applied magnetic field relative to 
 non-interacting case with periodic boundary condition, for the case $U=4$, $V=2$. 
 The conventional scenario (no flux) gives somewhat larger charge current than the scenario proposed here.
 The small kink in the $\pi-$ flux case near $J=0.7$ corresponds to the energy level crossing point in Fig. 6.
 }
 \label{figure2}
 \end{figure}

  Thus, for $U=4$ the range of $V$ where a spin current carrying ground state for
  an infinitesimal $\phi_s$  exists is restricted to
  $1.83\leq V\leq 2.05$.
  Similarly restricted parameter ranges where spin current in the ground state can develop for infinitesimal
  spin-orbit coupling exist for other $U$ values.

  The response of the system to an applied external magnetic field is similar to the conventional scenario for the case considered here.
  Figure 11 shows the charge current that develops for the case $U=4$, $V=2$ as function of $J$, relative to the
  current expected within Huckel theory (non-interacting electrons) in the presence of a small magnetic flux, of the same sign
  for both spin orientations. Both the conventional scenario with no spin-orbit flux and 
  the scenario with $\pi-$flux proposed here give a charge current, hence a diamagnetic susceptibility, that is almost constant as
  function of the interaction $J$ and not much smaller than the non-interacting value (which is $1$ in Fig. 11). The small glitch in the lower
  curve corresponds to the point where the energy level crossing occurs in Fig. 6.

    \section{spin-orbit coupling}
    If the ground state of the interacting Hamiltonian is degenerate (or nearly degenerate) in the way discussed in the previous section,
    spin-orbit coupling will split the degeneracy and give rise to two spin-current-carrying low-lying states.
    The lowest of the two states is shown schematically in Fig. 12.
             \begin{figure}
 \resizebox{7.5cm}{!}{\includegraphics[width=8cm]{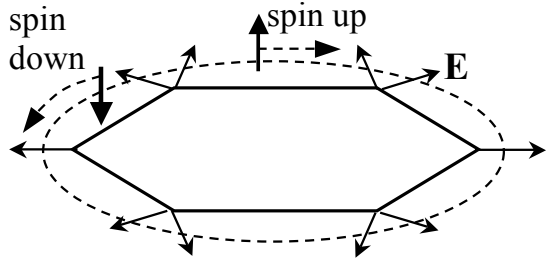}}
 \caption  { Electrons move in an outward-pointing electric field $\bf{E}$ originating in the charge transfer from the $C$ atom to the $C-H$ bond.
 The spin down (up) electron has lower energy when orbiting counterclockwise (clockwise), as shown in the figure,
 due to spin-orbit coupling. 
}
 \label{figure2}
 \end{figure}

 We can understand the physics as follows. 
  The electron in the spin current moves in an electric field pointing outwards, just like in an atom or in the superconductor
  according to the theory of hole superconductivity\cite{sm}. 
  The spin-orbit interaction Hamiltonian in electric field $\vec{E}$ is 
 \beq
 H_{so}=-\frac{e\hbar}{4m_e^2c^2}\vec{\sigma}\cdot(\vec{E}\times\vec{p})
 \eeq
 and it gives rise to an energy splitting 
 \beq
 \Delta E =\frac{e \hbar}{2m_e^2c^2} Ep
 \eeq
 between electrons with spin parallel and antiparallel to $\vec{E}\times\vec{p}$. Taking $p\sim\hbar/a$, with $a$ the 
 radius of the orbit (or equivalently the $C-C$ distance)
  \beq
 \Delta E =\frac{e \hbar^2}{2m_e^2c^2a} E
 \eeq
 The spin-orbit interaction   gives rise to the Aharonov-Casher vector potential\cite{ac}
 \beq
 \vec{A}_\sigma=\frac{1}{2e}\vec{\mu}\times\vec{E}
 \eeq
 with $\vec{\mu}=\mu_B\vec{\sigma}$, $\mu_B$ the Bohr magneton.
In going around the ring the electron acquires a phase shift
\beq
    \delta \theta_{AC} = \frac{e}{\hbar c}\oint A_\sigma\cdot dl =
    \frac{2\pi a e}{\hbar c}A_\sigma = \frac{\pi a e}{2m_e c^2} E
    \eeq
    or in terms of the energy splitting Eq. (17)
    \beq
     \delta \theta_{AC} =\frac{m_e}{\hbar^2}\pi a^2 \Delta E
     \eeq
    Assuming the energy splitting $\Delta E$ is given by the atomic spin-orbit splitting in C between $^2P_{3/2}$ and
    $^2P_{1/2}$ states, $\Delta E=8meV$, gives for the AC phase Eq. (20) $\delta \theta_{AC}=0.065$. The staggered flux
    introduced in Eq. (12) is then $1/6$th of this value, i.e.
    \beq
    \phi_s=0.0011
    \eeq
    close to the middle value of $\phi_s$ used for illustration in Fig. 5. For spin-orbit energy splitting a factor of $10$ larger,
    i.e. $ \Delta E\sim 80 meV$, the flux corresponds to the largest value shown in Fig. 5, where a rather large spin current
    occurs in a wide range of $J$ values.
    
    In summary, we propose that the   two lowest energy states of benzene carry a large spin current around the ring, as depicted
    schematically in Fig. 4 (a) and (b), and have a small energy separation of order $meV$. The ground state has
    spin current direction as shown in Fig. 12.

                 \begin{figure}
 \resizebox{8.5cm}{!}{\includegraphics[width=8cm]{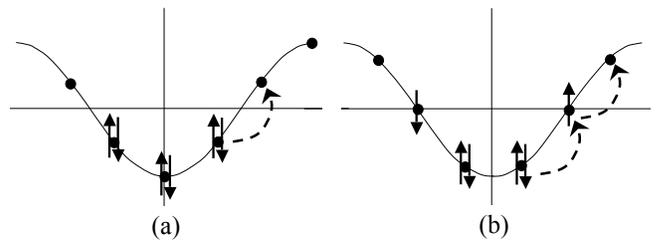}}
 \caption  { Allowed optical transitions in benzene in the conventional scenario ((a)) and in the scenario proposed here ((b))
 are indicated by the dashed lines.
}
 \label{figure2}
 \end{figure}
 
  \section{physical consequences}
  
  One immediate consequence of the scenario proposed here concerns optical properties and is illustrated in Fig. 13.
  As discussed by Platt\cite{platt}, allowed optical transitions in benzene require  change of one unit of angular momentum, 
  corresponding to change of the electron wave vector by $2\pi/6$.
  In the conventional scenario there is only one way to do this starting from the ground state, as shown schematically
  in Fig. 13 (a). In the scenario proposed here there are two distinct ways of exciting electrons to states differing in 
  wave vector by $2\pi/6$, as shown in Fig. 13 (b). The photon wavelengths associated with these two processes should be
  different because of the different effects of electron-electron interactions in the final states.

  Experimentally\cite{murrell} three bands are seen in the optical absorption spectrum of benzene, at wavelengths around $\lambda=2600A$,  $2050A$ and $1850A$. The one at 
  $2600$ is weak and has considerable fine structure, thus it is likely to arise from `forbidden transitions' involving
  interaction with vibrational degrees of freedom\cite{herzberg}. The other two bands are much stronger, and their interpretation
  has been controversial, because there is only one allowed transition in the conventional scenario (Fig. 13(a)). 
  There is general agreement that the transition at $\lambda=1850A$ is an allowed one because it is the strongest, and  the transition at $\lambda=2050A$ has been attributed to one or another forbidden transition. A variety of calculations of oscillator strength have been
  reported\cite{fb1,fb2,fb3,fb4} but it is not clear that these complicated calculations explain the experimental observations.  
  Instead, our scenario predicts two allowed transitions (Fig. 13(b)) and hence two strong absorption bands in the optical absorption of benzene, thus
offering the possibility of a more natural explanation for the observations. 

The ground state and low-lying energy state proposed here for benzene are neither pure singlet nor triplet, instead singlet and
triplet are strongly mixed, with nearly equal weights if $J$ is near the crossing point of singlet and triplet energy levels shown
in the figures.
It is possible that transitions from higher excited states to these states will provide a more natural explanation of the
phosphorescence properies of benzene than the conventional understanding, within which phosphorescence in benzene is
`doubly forbidden' and  
requires   vibronic interactions to play a key role\cite{phospho}.
          \begin{figure}
 \resizebox{8.5cm}{!}{\includegraphics[width=8cm]{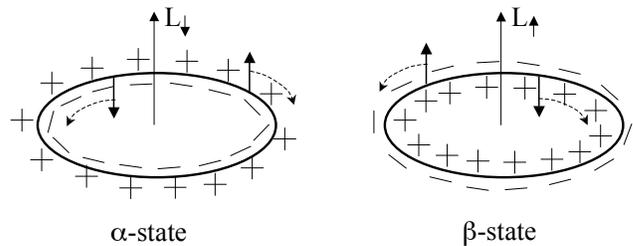}}
 \caption  { Dipolar charge distribution produced by spin current. $\alpha-$state: spin antiparallel to orbital angular momentum,
$\beta-$state: spin parallel to orbital angular momentum. $L_\sigma$ denotes the orbital angular momentum for spin $\sigma$
 and only the upward pointing $L_\sigma$   is shown.  The $\alpha-$state will generally be lower in energy due to
 spin-orbit coupling.
}
 \label{figure2}
 \end{figure}
 
The spin current proposed to exist in benzene will give rise to an electric field\cite{ss}.
A moving magnetic moment $\vec{\mu}$ is equivalent to an electric dipole moment 
 $\vec{p}$ in the
 laboratory frame
 \beq
 \vec{p}=\frac{\vec{v}}{c}\times\vec{\mu}
 \eeq
 and will give rise to an `effective' dipolar charge distribution depicted schematically in Fig. 14. 
This dipolar charge distribution
 generates a quadrupolar electric field around the aromatic ring. For a point  in the plane at distance $x>>a$ from the center of the ring
 \beq
 E(x) \sim \frac{6pa}{x^4}
 \eeq
 with $a$ the radius of the ring. At distance $x=2a$ of the center, the electric field is $E\sim p/a^3$.
 From Eq. (22) the dipole moment is $p\sim 77\mu V A^2$ for benzene, and the magnitude of the electric field at $x=2a$ is $E \sim 54 \mu V/A$.

  We suggest that the two low-lying states of the aromatic ring depicted in Fig. 14, which we call
  $\alpha$ and $\beta$ states,  are stable,
 because of the topological constraint that the `winding number' for electrons going around the ring
 is $1/2$, and that a large energy barrier exists for switching between $\alpha$ and $\beta$ states.
  If these states exist in aromatic rings they are likely to play a crucial role in biological matter
 where aromatic rings are ubiquitous. 
 
 How can an aromatic ring `switch' from one to the other low-lying state? 
The $\beta$ state becomes the lower energy state if the electric field that the electrons in the spin current `see' points
inward rather than outward, which could occur upon changes in the electrostatic field around the ring caused by
chemical reactions involving charge transfer, or transport of ions or electrons. 
A proton going through the interior of a benzene ring\cite{proton} may switch 
a $\beta$ to an $\alpha$ configuration due to the additional strong electric field pointing outward, and conversely an electron going through the
center of a benzene ring may switch the $\alpha$ to the $\beta$ state.

It is   tempting to speculate that the $\alpha$ and   $\beta$ states could act as quantum qubits in a topological quantum
computer\cite{qc}, namely the  brain. 
Electric fields originating in the spin current, of the magnitude estimated above, and resulting  electric potential  differences  may be involved in neural signal transmission.
Or, the $\alpha$ and $\beta$ states in aromatic rings in the brain may just be  storing one classical bit of information and play a key role
in memory storage\cite{ss}. Phase coherence of spin ring currents over large distances could play an important role in various aspects of brain activity such as
sleep.  
More generally, these stable dynamic low-lying states  of aromatic rings may  play a key role
in a wide variety of biological processes in living organisms\cite{aworld}.

As discussed earlier, the theory of hole superconductivity predicts that superconductors in the ground state  also carry a 
spin current and associated $\pi$ flux. The ground state of a small superconducting ring 
corresponds  to the $\alpha$ state in Fig. 14. It should be possible to switch such a ring to the $\beta$ state, 
perhaps through transient application of a large electrostatic field or a very large magnetic field. Such ring would exhibit a stable persistent spin current in 
opposite direction to that of the ground state (in the absence of external fields), and experimental detection of such a state would provide convincing
proof of the theory discussed here and in ref.\cite{sm}. The existence of such spin currents in superconducting
rings could be experimentally demonstrated
by detection of the quadrupolar electric fields generated by the dipolar charge distributions shown in Fig. 14, or by 
measuring the resulting force between small superconducting rings with spin currents, as discussed in ref.\cite{sm}.

  \section{discussion}
  
  Aromatic rings have $4n+2$ atoms, with $n$ integer. It is generally believed that such molecules have a singlet ground state
  described by delocalized electrons in a half-filled band\cite{pauling,haddon}, with a large energy gap to the first excited state.
  Alternatively, it has been proposed that a similar energy level structure and aromatic character can be described by 
  a valence bond state with localized electrons and coupling between electron spins\cite{vb}.  In this paper we have
  proposed instead a qualitatively different scenario: that the ground state of aromatic ring molecules is a mixture of singlet and
  triplet states that carries a spin current around the ring, and that another such state exists nearby in energy carrying spin current
  in the opposite direction. We have also proposed that the same is true for small superconducting rings
  made of ordinary superconductors.
  
  This scenario follows from the fundamental assumption that when an electron goes around a closed path in a coherent way it
  acquires a $\pi$ phase shift, and two rounds are needed to bring it back to the original state, like a point in a Moebius strip. In the
  aromatic ring, starting in a $p_z$ orbital the electron ends up in the $(-p_z)$ orbital after one round. We have justified this
  assumption by invoking the transformation properties of the spinor under rotation in quantum mechanics\cite{merzbacher},
  and by analogy with the phase behavior predicted by the theory of hole superconductivity,
stemming from the finding that electrons in a Cooper pair carry orbital angular momentum $\hbar/2$\cite{sm}.

Within this assumption, and using a Hubbard-like Hamiltonian including nearest neighbor exchange to model
the benzene ring, we have found
reasonable parameter ranges in the interacting system where a spin-current-carrying ground state and low-lying excited state exists. The same is likely to be
true for other aromatic ring structures using the same model.
We have argued that our scenario provides a more natural explanation for the optical properties of benzene
as well as for the existence of ring currents evidenced by its large diamagnetic susceptibility and proton NMR shifts.

Whether our proposal is true or false can be decided experimentally. It should be possible to make aromatic rings and
small superconducting rings in both predicted low-lying spin current states and study their properties.

If our proposal is true it is likely to yield fundamental insights into the properties of biological matter, as well as on
quantum mechanics itself, changing the currently accepted understanding in fundamental ways. It is also likely to
open up possibilities of qualitatively new technological advances such as quantum computation.
 
 \acknowledgements
 The author is grateful to Congjun Wu for stimulating discussions.

\end{document}